# Anisotropic hole drift velocity in 4H-SiC

Jackelinne L. Vasconcelos[a], Clóves G. Rodrigues[a],[*], Roberto Luzzi[b]

[a] *School of Exact Sciences and Computing, Pontifical Catholic University of Goiás, CP 86, 74605-010 Goiânia, Goiás, Brazil*
[b] *Condensed Matter Physics Department, Institute of Physics "Gleb Wataghin" State University of Campinas-Unicamp, 13083-859 Campinas, SP, Brazil*

## ABSTRACT

A theoretical study on the nonlinear transport of holes in the transient and steady state of p-doped 4H-SiC under the influence of high electric fields is presented. It is based on a nonlinear quantum kinetic theory which provides a clear description of the dissipative phenomena that are evolving in the system. The hole drift velocity and the nonequilibrium temperature are obtained, and their dependence on the electric field strength is derived and analyzed.

## 1. Introduction

The intrinsic properties of the wide band gap semiconductors, specifically the wide band gap energy that enables higher junction operating temperatures, high saturated electron drift velocity, small dielectric constant, and in the case of SiC a relatively high thermal conductivity, make these materials particularly attractive for high power device applications. The most attractive candidate materials for high power device applications are GaN and some of the various polytypes of SiC [1–5]. In both of these materials, the breakdown electric field strengths are expected to be about four times larger than in either silicon or GaAs. In addition to the similar properties of GaN (large energy band gap, high carrier saturation drift velocity, and relatively small dielectric constant) SiC has a significantly higher thermal conductivity: a high thermal conductivity is necessary in order to overcome device-heating effects which is particularly important in power amplifiers [1–5].

Silicon carbide can form in many distinct crystal structures (know as polytypes), with some of the most common being 3C-SiC (cubic), 4H-SiC (hexagonal) and 6H-SiC (hexagonal) [6]. Hexagonal SiC polytypes, 4H-SiC and 6H-SiC, commonly used for the fabrication of devices, have anisotropic transport properties, but the degree and characteristics anisotropy are different. The anisotropy is associated with the hexagonal symmetry, which causes the hole effective masses to differ significantly between orientations parallel to the *c*-axis (which we call $m^*_{h\|}$), and those along the basal plane perpendicular to the *c*-axis (which we call $m^*_{h\perp}$). Among the several polytypes in SiC, 4H-SiC has been recognized as the most attractive material for electronic devices in high-power, high-frequency, and high-temperature operations because of its wider band gap and higher electron mobility than other polytypes [7–12].

The optical and transport properties of semiconductors have been studied by using Nonequilibrium Green's Functions Techniques, Monte Carlo simulation, balance equation, etc. Here, we use the NESEF (Non Equilibrium Statistical Ensemble Formalism) [13–23]. The NESEF is practical and efficient in the study of the transport and optical properties of semiconductors [24–29]. More specifically, we use a Nonequilibrium Quantum Kinetic Theory derived from NESEF [30].

In this work NESEF has been used for the study of bulk nonlinear transport properties in 4H-SiC when the transport direction is along the *c*-axis, or when the transport direction is in the plane perpendicular to it. The hole drift velocity and the nonequilibrium temperature in the transient and steady state are obtained, and their dependence on the electric field is analyzed. To develop high performance electronic devices, beyond optimizing the fabrication steps, a good knowledge of the transport properties is required. As an example, the carrier mobility is a very important property, affecting the device performances and, hence, can be considered as a figure of merit of the microscopic quality of epilayers. Depending on the particular device, one is interested in the carrier mobility either parallel to the *c*-axis or perpendicular to it. In particular, the carrier mobility along the *c*-axis is extremely important in vertical power devices such as Schottky diodes.

## 2. Theoretical background

The Hamiltonian of the p-doped semiconductor is taken as composed of: the energy of the free phonons, the energy of the free holes, the interaction of the holes with the constant electric field, the interaction of the holes with the impurities, the hole-phonon interaction, the anharmonic interaction of the TO-phonons with AC-phonons, the anharmonic interaction of the LO-phonons with AC-phonons, the interaction of the AC-phonons with the thermal bath (external thermal reservoir at temperature $T_0$). In Section III we use the parameters corresponding to the semiconductor 4H-SiC for numerical calculations.

[*] Corresponding author.
  *E-mail address:* cloves@pucgoias.edu.br

**Table 1**
Parameters of the 4H-SiC [6,34,35].

| Parameter | Value |
|---|---|
| Hole effective mass for $\mathcal{E}_\perp$: $m^*_{h\perp}$ | $0.59 m_0$ |
| Hole effective mass for $\mathcal{E}_\parallel$: $m^*_{h\parallel}$ | $1.60 m_0$ |
| Lattice constant $a$ | 3.073 Å |
| Lattice constant $c$ | 10.053 Å |
| Longit. optical phonon energy $\hbar\omega_{LO}$ | 120 meV |
| Transv. optical phonon energy $\hbar\omega_{TO}$ | 98 meV |
| Mass density | 3.12 g/cm$^3$ |
| Longitudinal sound velocity $v_{sl}$ | $1.355 \times 10^6$ cm/s |
| Transverse sound velocity $v_{st}$ | $7.5 \times 10^5$ cm/s |
| Acoustic deformation potential $E_1$ | 6.5 eV |
| Optical deformation potential $D_{op}$ | $3.2 \times 10^9$ eV/cm |
| Low frequency dielectric constant $\epsilon_0$ | 10.0 |
| High frequency dielectric constant $\epsilon_\infty$ | 6.7 |

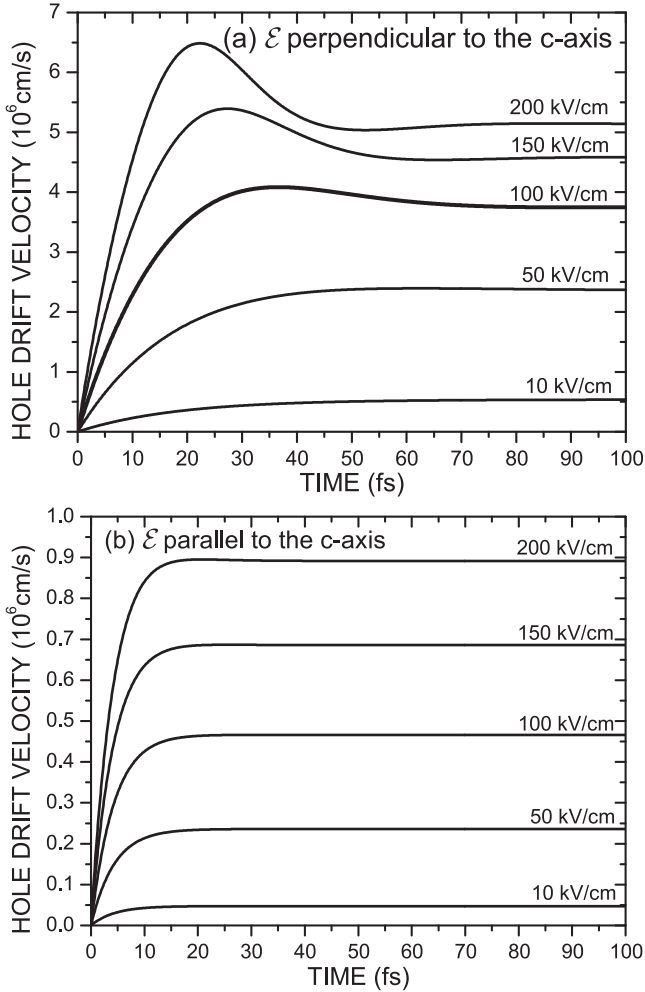

**Fig. 1.** Time evolution towards the steady state of the hole drift velocity, $v_h(t)$, of p-doped 4H-SiC for five values of the electric field intensity. (a) $\mathcal{E}$ perpendicular to the $c$-axis, and (b) $\mathcal{E}$ parallel to the $c$-axis.

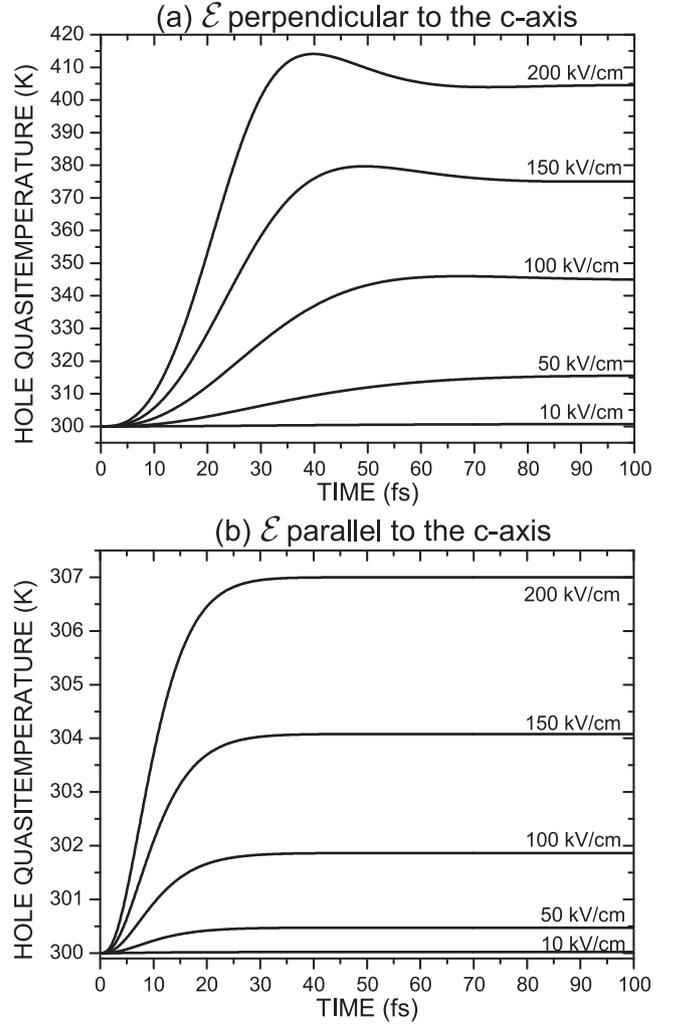

**Fig. 2.** Time evolution towards the steady state of the nonequilibrium temperature of holes, $T_h^*(t)$, of p-doped 4H-SiC for five values of the electric field intensity. (a) $\mathcal{E}$ perpendicular to the $c$-axis, and (b) $\mathcal{E}$ parallel to the $c$-axis.

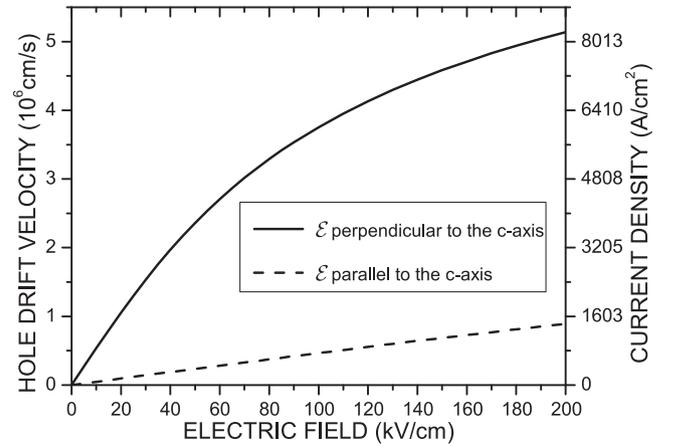

**Fig. 3.** Dependence on the electric field of the hole drift velocity, $v_h(\mathcal{E})$, in the steady state of p-doped 4H-SiC. Solid line: $\mathcal{E}$ perpendicular to the $c$-axis. Dashed line: $\mathcal{E}$ parallel to the $c$-axis.

The nonequilibrium thermodynamic state of the system is very well characterized for the set of *basic variables*:

$$\{E_h(t), N, \mathbf{P}(t), E_{LO}(t), E_{TO}(t), E_{AC}\}, \qquad (1)$$

that is: the energy of the holes $E_h(t)$; the number $N$ of the holes; the linear momentum $\mathbf{P}(t)$ of the holes; the energy of the LO-phonons $E_{LO}(t)$; the energy of the TO-phonons $E_{TO}(t)$; and the energy of the AC-phonons

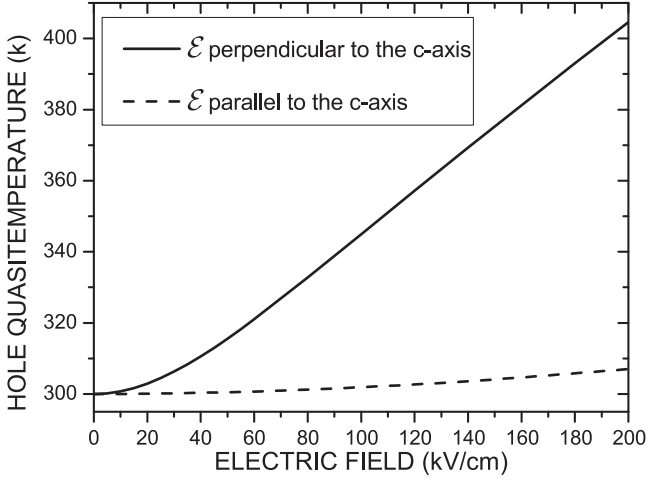

**Fig. 4.** Dependence on the electric field of the nonequilibrium temperature of holes, $T_h^*(\mathcal{E})$, in the steady state of p-doped 4H-SiC. Solid line: $\mathcal{E}$ perpendicular to the *c*-axis. Dashed line: $\mathcal{E}$ parallel to the *c*-axis.

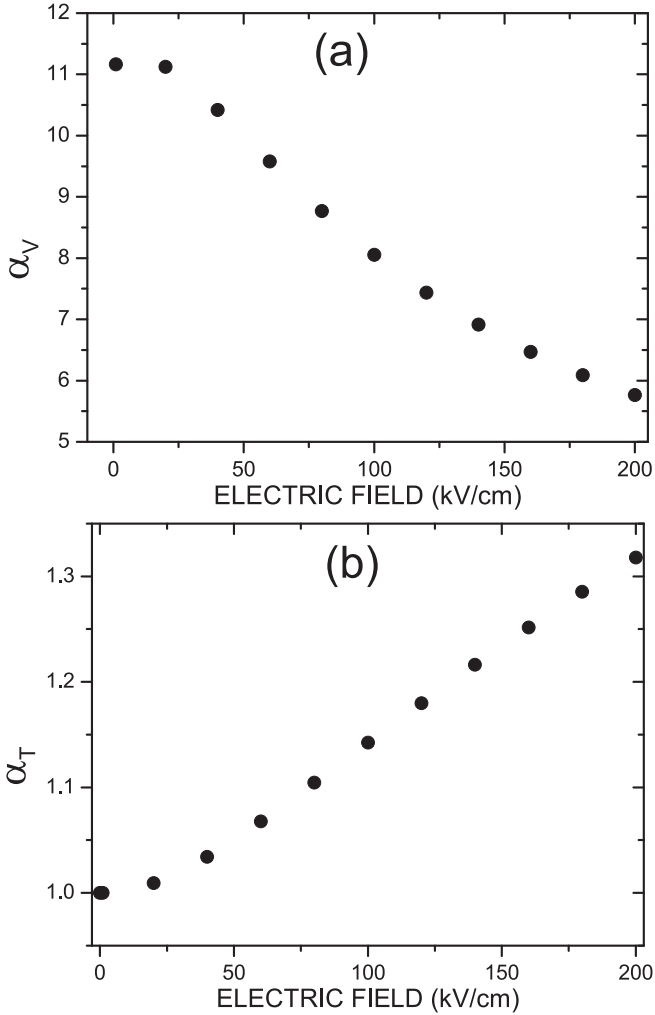

**Fig. 5.** (a) The ratio $\alpha_v = v(\mathcal{E}_\perp)/v(\mathcal{E}_\parallel)$, and (b) the ratio $\alpha_T = T_h^*(\mathcal{E}_\perp)/T_h^*(\mathcal{E}_\parallel)$.

$E_{AC}$.

The *nonequilibrium thermodynamic variables* associated with variables of set (1), according to the NESEF [16–18,31], are

$$\{\beta_h^*(t), -\beta_h^*(t)\mu_h^*(t), -\beta_h^*(t)\mathbf{v}(t), \beta_{LO}^*(t), \beta_{TO}^*(t), \beta_{AC}^*(t)\}. \quad (2)$$

where $\mathbf{v}(t)$ is the drift velocity of the holes, $\mu_h^*(t)$ is the quasi-chemical potential and

$$\beta_h^*(t) = \frac{1}{k_B T_h^*(t)}, \quad (3a)$$

$$\beta_{LO}^*(t) = \frac{1}{k_B T_{LO}^*(t)}, \quad (3b)$$

$$\beta_{TO}^*(t) = \frac{1}{k_B T_{TO}^*(t)}, \quad (3c)$$

$$\beta_{AC}^*(t) = \frac{1}{k_B T_{AC}^*(t)}, \quad (3d)$$

where $k_B$ is Boltzmann constant, $T_h^*(t)$ is the nonequilibrium temperature of the holes, $T_{LO}^*(t)$ is the nonequilibrium temperature of the LO-phonons, $T_{TO}^*(t)$ is the nonequilibrium temperature of the TO-phonons and $T_{AC}^*(t)$ is the nonequilibrium temperature of the AC-phonons [31–33].

By using the Nonlinear Quantum Kinetic Theory based on the NESEF, we obtain the equations of evolution for the basic macrovariables [30]:

$$\frac{d\mathbf{P}(t)}{dt} = -nVe\mathcal{E} + \mathbf{J}_{\mathbf{P}_{ph}}(t) + \mathbf{J}_{\mathbf{P}_{imp}}(t), \quad (4)$$

$$\frac{dE_h(t)}{dt} = -\frac{e}{m_h^*}\mathcal{E}\cdot\mathbf{P}(t) + J_{E,ph}(t), \quad (5)$$

$$\frac{dE_{LO}(t)}{dt} = J_{LO}(t) - J_{LO,an}(t), \quad (6)$$

$$\frac{dE_{TO}(t)}{dt} = J_{TO}(t) - J_{TO,an}(t), \quad (7)$$

$$\frac{dE_{AC}(t)}{dt} = J_{AC}(t) + J_{LO,an}(t) + J_{TO,an}(t) - J_{AC,dif}(t), \quad (8)$$

where $n$ is the concentration of holes fixed by doping (and consequently $dn/dt = 0$); $E_h(t)$ is the energy of the holes; $\mathbf{P}(t)$ is the linear momentum of the holes; $E_{LO}(t)$ is the energy of the longitudinal optical phonons that interact with the holes via optical deformation potential and Fröhlich potential; $E_{TO}(t)$ is the energy of the transverse optical phonons that interact with the holes via optical deformation potential; $E_{AC}(t)$ is the energy of the acoustic phonons that interact with the holes via acoustic deformation potential; $\mathcal{E}$ is the constant electric field applied perpendicular ($\mathcal{E}_\perp$) or parallel ($\mathcal{E}_\parallel$) to the *c*-axis direction. Moreover, $m_h^* = m_{h\parallel}^*$ when the constant electric field is applied parallel to the *c*-axis, or $m_h^* = m_{h\perp}^*$ when the constant electric field is applied perpendicular to the *c*-axis. These values and other parameters for the study of transport phenomena in 4H-SiC are shown in Table 1. We notice that in Table 1 $m_0$ stands for the free electron mass.

In Eq. (4), the first right hand term ($-nVe\mathcal{E}$) is the driving force created by the constant electric field applied. The second term, $\mathbf{J}_{\mathbf{P}_{ph}}(t)$, is the rate of hole momentum transfer due to interaction with the LO, TO and AC phonons. The third term, $\mathbf{J}_{\mathbf{P}_{imp}}(t)$, is the rate of hole momentum transfer due to interaction with the ionized impurities [36].

In Eq. (5), the first right hand term ($-e\mathcal{E}\cdot\mathbf{P}(t)/m_h^*$) is the rate of energy transferred from the constant electric field applied to the holes, and the second term, $J_{E,ph}(t)$, is the transfer of the energy of the holes to the LO, TO and AC phonons.

In Eqs. (6)–(8), the first right hand term accounts for the rate of change of the energy of the LO, TO or AC phonons, respectively, due to interaction with the holes, that is, the gain of energy transferred to the phonons from the hot holes. Like this, the sum of contributions $J_{LO}(t)$, $J_{TO}(t)$ and $J_{AC}(t)$ is equal to the last term in Eq. (5), $J_{E,ph}(t)$, but with change of sign. The second term in Eq. (6), $J_{LO,an}(t)$, is the rate of transfer of energy from the LO-phonons to the AC-phonons via anharmonic interaction. We noticed that the therm $J_{LO,an}(t)$ is the same, however with different sign in Eqs. (6) and (8). The second term in Eq.

**Table 2**
Comparative values for the mobility in p-doped 4H-SiC.

| Reference | Method | Mobility (cm$^2$/V.s) |
| --- | --- | --- |
| Ref. [39] | Theoretical | 16.0 |
| Ref. [40] | Experimental | 45.2 |
| this work | theoretical | 54.1 |
| Ref. [41] | Experimental | 56.7 |
| Ref. [42] | Experimental | 67.5 |
| Ref. [43] | Experimental | 68.9 |
| Ref. [10] | Experimental | 74.7 |
| Ref. [44] | Experimental | 96.3 |
| Ref. [45] | Experimental | 100.1 |
| Ref. [46] | Simulation | 102.6 |
| Ref. [47] | Simulation | 116.1 |

(7), $J_{\text{TO},an}(t)$, is the rate of transfer of energy from the TO-phonons to the AC-phonons via anharmonic interaction. Analogously, the contribution $J_{\text{TO},an}(t)$ is the same, however with different sign in Eqs. (7) and (8). Shutting down our analysis of Eqs. (4)–(8), we noticed that the last term in Eq. (8), $J_{\text{AC},dif}(t)$, is the diffusion of heat from the AC-phonons to the thermal bath at temperature $T_0$.

We notice that the detailed expressions for the collision operators: $\mathbf{J}_\mathbf{P}(t)$, $J_{E,ph}(t)$, $J_{\text{LO}}(t)$, $J_{\text{LO},an}(t)$, $J_{\text{TO}}(t)$, $J_{\text{TO},an}(t)$, $J_{\text{AC}}(t)$, and $J_{\text{AC},dif}(t)$ are given in Ref. [30], which are for electrons, appropriately adapted for holes in this work.

We emphasize that to close the system of equations of evolution we must establish the relationship between the *basic variables*, Eq. (1), and the *nonequilibrium thermodynamic variables*, Eq. (2). These relationships are:

$$E_h(t) = \sum_\mathbf{k} \epsilon_\mathbf{k} f_\mathbf{k}(t) = nV\left[\frac{3}{2}k_B T_h^*(t) + \frac{1}{2}m_h^* v(t)^2\right], \quad (9)$$

$$\mathbf{P}(t) = \sum_\mathbf{k} \hbar\mathbf{k} f_\mathbf{k}(t) = nV m_h^* \mathbf{v}(t), \quad (10)$$

$$E_{\text{LO}}(t) = \sum_\mathbf{q} \hbar\omega_{\text{LO}} \nu_{\text{LO}}(t), \quad (11)$$

$$E_{\text{TO}}(t) = \sum_\mathbf{q} \hbar\omega_{\text{TO}} \nu_{\text{TO}}(t), \quad (12)$$

$$E_{\text{AC}}(t) = \sum_\mathbf{q} \hbar\omega_{\text{AC}} \nu_{\text{AC}}(t), \quad (13)$$

which are known as "*nonequilibrium thermodynamic equations of state*" [16–18,30,31].

In a closed calculation, by using the NESEF, we obtain to phonon populations

$$\nu_{\text{LO}}(t) = \frac{1}{e^{\beta_{\text{LO}}^*(t)\hbar\omega_{\text{LO}}} - 1}, \quad (14)$$

$$\nu_{\text{TO}}(t) = \frac{1}{e^{\beta_{\text{TO}}^*(t)\hbar\omega_{\text{TO}}} - 1}, \quad (15)$$

$$\nu_{\text{AC}}(t) = \frac{1}{e^{\beta_{\text{AC}}^*(t)\hbar\omega_{\text{AC}}} - 1}, \quad (16)$$

and for to hole populations (in the nondegenerate state)

$$f_\mathbf{k}(t) = \exp\{-\beta_h^*(t)[\epsilon_\mathbf{k} - \hbar\mathbf{k}\cdot\mathbf{v}(t) - \mu_h^*(t)]\}, \quad (17)$$

where

$$\mu_h^*(t) = \frac{1}{\beta_h^*(t)}\ln\left\{\frac{4n\hbar^3}{e^{\beta_h^*(t)m_h^* v(t)^2/2}}\sqrt{\left(\frac{\pi\beta_h^*(t)}{2m_h^*}\right)^3}\right\} \quad (18)$$

We emphasize that we take the "Einstein model" dispersionless frequency relation for LO and TO phonons and the "Debye model" dispersionless frequency relation for AC phonons.

The hole energy dispersion relation $\epsilon_\mathbf{k}$ (three sub-bands) are highly non-parabolic – due to the small spin-orbit splitting –, what implies in effective hole masses becoming strongly **k**-dependent. However, we shall restrict the situation to the conditions of low densities and weak to moderate electric field intensities so that the parabolic band approximation is acceptable, i.e., we take $\epsilon_\mathbf{k} = \hbar^2\mathbf{k}\cdot m_h^*\mathbf{k}/2$.

## 3. Results and conclusions

The time evolution of the basic intensive nonequilibrium thermodynamic variables are obtained after numerically solve the set of coupled nonlinear integro-differential equations: Eqs. (4)–(8). We assumed that the reservoir temperature is $T_0 = 300$ K, and that the dopant concentration $n_I$ is equal to the hole concentration $n$, with $n = n_I = 10^{16}$ cm$^{-3}$. Moreover, we noticed that the electric fields applied to 4H-SiC (applied parallel to the *c*-axis or applied perpendicular to the *c*-axis) is restricted to be smaller than 200 kV/cm since intervalley scattering is not considered in this work. We have used the parameters for 4H-SiC listed in Table 1 to perform the numerical calculations.

The time evolution of the hole drift velocity, $v_h(t)$, and the hole nonequilibrium temperature, $T_h^*(t)$, towards the steady state, for five values of the electric field intensity, is depicted in Figs. 1 and 2, respectively. In Figs. 1 and 2, the upper figure is for the $\mathscr{E}$ applied perpendicular to the *c*-axis and the lower figure is for the $\mathscr{E}$ applied parallel to the *c*-axis. It can be noticed that the time for the holes to attain the steady state is in this paper very approximately the same obtained previously by other authors using different descriptions of the transient transport phenomena [34,37]. Figs. 1 and 2 show the existence of an overshoot in both the hole nonequilibrium temperature and the hole drift velocity. The overshoot occurs when the carrier relaxation rate of energy is less than the carrier relaxation rate of momentum during the evolution dynamics of the macroscopic state of the system [38].

Fig. 3 shows the dependence of the hole drift velocity, in the steady state, with the electric field strength, and Fig. 4 shows the dependence on the electric field strength of the nonequilibrium temperatures of holes in the steady state of 4H-SiC. In Figs. 3 and 4, the upper figure is for the $\mathscr{E}$ applied perpendicular to the *c*-axis and the lower figure is for the $\mathscr{E}$ applied parallel to the *c*-axis. Looking at Fig. 3, it can be noticed that the lower drift velocity corresponds to $\mathscr{E}$ applied parallel to the *c*-axis. This is a consequence that the holes have a greater effective mass in the direction parallel to the *c*-axis than in the direction perpendicular to the *c*-axis (see Table 1).

The Fig. 5(a) shows the ratio $\alpha_v = v(\mathscr{E}_\perp)/v(\mathscr{E}_\parallel)$, that is, the ratio between the hole drift velocity with the $\mathscr{E}$ applied perpendicular to the *c*-axis ($\mathscr{E}_\perp$) and the hole drift velocity with the $\mathscr{E}$ applied parallel to the *c*-axis ($\mathscr{E}_\parallel$). Moreover, the Fig. 5(b) shows the ratio $\alpha_T = T_h^*(\mathscr{E}_\perp)/T_h^*(\mathscr{E}_\parallel)$, that is, the ratio between the hole nonequilibrium temperature with the $\mathscr{E}$ applied perpendicular to the *c*-axis ($\mathscr{E}_\perp$) and the hole nonequilibrium temperature with the $\mathscr{E}$ applied parallel to the *c*-axis ($\mathscr{E}_\parallel$). Inspection of Figs. 5(a) and 5(b) tells us that the difference between $v(\mathscr{E}_\perp)$ and $v(\mathscr{E}_\parallel)$ decreases with increasing of the electric field intensity, and the difference between $T_h^*(\mathscr{E}_\perp)$ and $T_h^*(\mathscr{E}_\parallel)$ increases with increasing of the electric field intensity.

An important characteristic regarding the transport of electric charge in semiconductors is the carrier mobility, defined as $\mathscr{M} = v/\mathscr{E}$, where $v$ is the carrier drift velocity and $\mathscr{E}$ is the electric field intensity. Table 2 shows several hole mobility values obtained by experimental and theoretical methods for the p-doped 4H-SiC. All values given in Table 2 are for low electric fields, $T_0 = 300$ K and a hole concentration $<3 \times 10^{17}$ cm$^{-3}$. It is noted that our results are compatible with the experimental results of Sylvie et al. [40] and Kolaklieva et al. [41].

In conclusion, we have presented a study on the transient (and steady state) transport in p-doping 4H-SiC using a Nonequilibrium Quantum Kinetic Theories derived from the Non Equilibrium Statistical Ensemble Formalism (NESEF). A good knowledge of the transport properties is required to develop high performance electronic devices. Depending on the particular device, one is interested in the carrier mobility either parallel to the *c*-axis or perpendicular to it.

## References

[1] K. Takahashi, A. Yoshikawa, A. Sandhu, Wide Bandgap Semiconductors, Springer, Berlin, Germany, 2006.
[2] T. Kimoto, Material science and device physics in SiC technology for high-voltage power devices, Jpn. J. Appl. Phys. 54 (2015) 040103, https://doi.org/10.7567/JJAP.54.040103.
[3] T.K. Gachovska, J.L. Hudgins, SiC and GaN power semiconductor devices, Chapter 5, in: Muhammad H. Rashid (Ed.), Power Electronics Handbook, fourth ed., Butterworth-Heinemann, Oxford, UK, 2018, , https://doi.org/10.1016/C2016-0-00847-1.
[4] F. Roccaforte, P. Fiorenza, G. Greco, R. Nigro, F. Giannazzo, F. Iucolano, M. Saggio, Emerging trends in wide band gap semiconductors (SiC and GaN) technology for power devices, Microelectron. Eng. 187–188 (2018) 66–77, https://doi.org/10.1016/j.mee.2017.11.021.
[5] M. Hebali, D. Berbara, M. Benzohra, D. Chalabi, A. Saïdane, A comparative study on electrical characteristics of MOS (Si0.5Ge0.5) and MOS (4H-SiC) transistors in 130nm technology with BSIM3v3 model, Int. J Adv. Comput. Electron. Eng. 3 (9) (2018) 1–6.
[6] C. Persson, A.F. Silva, Electronic properties of intrinsic and heavily doped 3C-, nH-SiC (n = 2, 4, 6) and III-N (III = B, Al, Ga, In), in: M. Henini, M. Razeghi (Eds.), Optoelectronic Devices: III-Nitrides, Elsevier, London, UK, 2005, pp. 479–559 vol. 1, Chapter 17.
[7] S. Contreras, L. Konczewicz, R. Arvinte, H. Peyre, T. Chassagne, M. Zielinski, S. Juillaguet, Electrical transport properties of p-type 4H-SiC, Phys. Status Solidi A 214 (4) (2017) 1600679, https://doi.org/10.1002/pssa.201600679.
[8] A.R. Chowdhury, J.C. Dickens, A.A. Neuber, R.P. Joshi, Assessing the role of trap-to-band impact ionization and hole transport on the dark currents of 4H-SiC photo-conductive switches containing deep defects, J. Appl. Phys. 120 (2016) 245705, , https://doi.org/10.1063/1.4972968.
[9] P.A. Ivanov, A.S. Potapov, T.P. Samsonova, I.V. Grekhov, Electric-field dependence of electron drift velocity in 4H-SiC, Solid-State Electron. 123 (2016) 15–18, https://doi.org/10.1016/j.sse.2016.05.010.
[10] H. Fujihara, J. Suda, T. Kimoto, Electrical properties of n- and p-type 4H-SiC formed by ion implantation into high-purity semi-insulating substrates, Jpn. J. Appl. Phys. 56 (2017) 070306, , https://doi.org/10.7567/JJAP.56.070306.
[11] M. Cabello, V. Soler, G. Rius, J. Montserrat, J. Rebollo, P. Godignon, Advanced processing for mobility improvement in 4H-SiC MOSFETs: a review, Mat. Sci. Semicon. Proc. 78 (2018) 22–31, https://doi.org/10.1016/j.mssp.2017.10.030.
[12] L. Lina, J. Huanga, W. Yub, H. Taoc, L. Zhua, P. Wanga, Z. Zhanga, J. Zhang, Electronic structures and magnetic properties of (Ni, Al) co-doped 4H-SiC: a first-principles study, Comp. Mater. Sci. 155 (2018) 169–174, https://doi.org/10.1016/j.commatsci.2018.08.048.
[13] D.N. Zubarev, Nonequilibrium Statistical Thermodynamics, Consultants Bureau, New York, USA, 1974.
[14] D.N. Zubarev, V. Morozov, G. Röpke, Statistical Mechanics of Nonequilibrium Processes, Academie Verlag - Wiley VCH, Berlin, Germany, 1997 vols. 1 and 2.
[15] A.I. Akhiezer, S.V. Peletminskii, Methods of Statistical Physics, Pergamon, Oxford, UK, 1981.
[16] R. Luzzi, A.R. Vasconcellos, J.G. Ramos, Predictive Statistical Mechanics: A Nonequilibrium Statistical Ensemble Formalism, Kluwer Academic, Dordrecht, The Netherlands, 2002.
[17] R. Luzzi, A.R. Vasconcellos, On the nonequilibrium statistical operator method, Fortsch. Phys./Prog. Phys. 38 (11) (1990) 887–922, https://doi.org/10.1002/prop.2190381104.
[18] J.G. Ramos, A.R. Vasconcellos, R. Luzzi, A classical approach in predictive statistical mechanics: a generalized Boltzmann formalism, Fortsch. Phys./Prog. Phys. 43 (4) (1995) 265–300, https://doi.org/10.1002/prop.2190430402.
[19] L. Lauck, A.R. Vasconcellos, R. Luzzi, A nonlinear quantum transport theory, Physica A 168 (2) (1990) 789–819, https://doi.org/10.1016/0378-4371(90)90031-M.
[20] R. Luzzi, A.R. Vasconcellos, J.G. Ramos, C.G. Rodrigues, Statistical irreversible thermodynamics in the framework of Zubarev's nonequilibrium statistical operator method, Theor. Math. Phys. 194 (1) (2018) 4–29, https://doi.org/10.1134/S0040577918010038.
[21] A.L. Kuzemsky, Theory of transport processes and the method of the nonequilibrium statistical operator, Int. J. Mod. Phys. B 21 (17) (2007) 2821–2949, https://doi.org/10.1142/S0217979207037417.
[22] A.L. Kuzemsky, Nonequilibrium statistical operator method and generalized kinetic equations, Theor. Math. Phys. 194 (1) (2018) 30–56, https://doi.org/10.1134/S004057791801004X.
[23] V.G. Morozov, Memory effects and nonequilibrium correlations in the dynamics of open quantum systems, Theor. Math. Phys. 194 (1) (2018) 105–113, https://doi.org/10.1134/S0040577918010075.
[24] C.G. Rodrigues, A.R. Vasconcellos, R. Luzzi, Ultrafast relaxation kinetics of photo-injected plasma in III-nitrides, J. Phys. D Appl. Phys. 38 (2005) 3584–3589, https://doi.org/10.1088/0022-3727/38/19/007.
[25] C.G. Rodrigues, A.R. Vasconcellos, R. Luzzi, Non-linear electron mobility in n-doped III-nitrides, Braz. J. Phys. 36 (2A) (2006) 255–257 URL: http://www.sbfisica.org.br/bjp/files/v36_255.pdf.
[26] C.G. Rodrigues, A.R. Vasconcellos, R. Luzzi, Nonlinear transport in n-III-nitrides: selective amplification and emission of coherent LO phonons, Solid State Commun. 140 (2006) 135–140, https://doi.org/10.1016/j.ssc.2006.08.015.
[27] C.G. Rodrigues, A.R. Vasconcellos, R. Luzzi, Optical properties of III-Nitrides under electric fields, Eur. Phys. J. B 72 (2009) 67–75, https://doi.org/10.1140/epjb/e2009-00332-y.
[28] C.G. Rodrigues, A.R. Vasconcellos, R. Luzzi, Evolution kinetics of nonequilibrium longitudinal-optical phonons generated by drifting electrons in III-nitrides: longitudinal-optical-phonon resonance, J. Appl. Phys. 108 (2010) 033716, , https://doi.org/10.1063/1.3462501.
[29] C.G. Rodrigues, A.A.P. Silva, C.A.B. Silva, A.R. Vasconcellos, J.G. Ramos, R. Luzzi, The role of nonequilibrium thermo-mechanical statistics in modern technologies and industrial processes: an overview, Braz. J. Phys. 40 (1) (2010) 63–91 URL: http://www.sbfisica.org.br/bjp/files/v40_63.pdf.
[30] C.G. Rodrigues, A.R. Vasconcellos, R. Luzzi, A kinetic theory for nonlinear quantum transport, Transp. Theor. Stat. Phys. 29 (2000) 733–757, https://doi.org/10.1080/00411450008200000.
[31] R. Luzzi, A.R. Vasconcellos, J.G. Ramos, Statistical Foundations of Irreversible Thermodynamics, Teubner-BertelsmannSpringer, Stuttgart, Germany, 2000.
[32] D. Jou, J. Casas-Vazquez, Temperature in non-equilibrium states: a review of open problems and current proposals, Rep. Prog. Phys. 66 (11) (2003) 1937–2024, https://doi.org/10.1088/0034-4885/66/11/R03.
[33] R. Luzzi, A.R. Vasconcellos, J. Casas-Vazquez, D. Jou, Thermodynamic variables in the context of a nonequilibrium statistical ensemble approach, J. Chem. Phys. 107 (1997) 7383–7398, https://doi.org/10.1063/1.474976.
[34] M.Z.S. Flores, F.F. Maia, V.N. Freire, J.A.P. Costa, E.F. Silva, Band structure anisotropy effects on the hole transport transient in 4H-SiC, Microelectron. J. 34 (2003) 717–719, https://doi.org/10.1016/S0026-2692(03)00110-1.
[35] E. Bellotti, H.E. Nilsson, K.F. Brennan, P.P. Ruden, R. Trew, Monte Carlo calculation of hole initiated impact ionization in 4H phase SiC, J. Appl. Phys. 87 (2000) 3864–3871, https://doi.org/10.1063/1.372426.
[36] B.K. Ridley, Reconciliation of the Conwell-Weisskopf and Brooks-Herring formulae for charged-impurity scattering in semiconductors: third-body interference, J. Phys. C: Solid State Physics 10 (1977) 1589–1593.
[37] H.E. Nilsson, A. Martinez, U. Sannemo, M. Hjelm, E. Bellotti, K. Brennan, Monte Carlo simulation of high field hole transport in 4H-SiC including band to band tunneling and optical interband, Physica B 314 (2002) 68–71, https://doi.org/10.1016/S0921-4526(01)01356-4.
[38] C.G. Rodrigues, A.R. Vasconcellos, R. Luzzi, V.N. Freire, Transient transport in III-nitrides: interplay of momentum and energy relaxation times, J. Phys.: Condens. Matter 19 (2007) 346214, , https://doi.org/10.1088/0953-8984/19/34/346214.
[39] A. Arvanitopoulosa, M. Antonioub, S. Perkinsa, M.R. Jenningsc, M.B. Guadasd, K.N. Gyftakise, N. Lophitis, On the suitability of 3C-Silicon Carbide as an alternative to 4H-Silicon Carbide for power diodes, IEEE T. Ind. Appl. (2019), https://doi.org/10.1109/TIA.2019.2911872 (in press).
[40] C. Sylvie, K. Leszek, K. Pawel, A. Roxana, P. Hervé, C. Thierry, Z. Marcin, K. Maria, J. Sandrine, Z. Konstantinos, Electrical transport properties of highly aluminum doped p-type 4H-SiC, Mater. Sci. Forum 858 (2015) 249–252, https://doi.org/10.4028/www.scientific.net/MSF.858.249.
[41] L. Kasamakova-Kolaklieva, L. Storasta, I.G. Ivanov, B. Magnusson, S. Contreras, C. Consejo, J. Pernot, M. Zielinski, E. Janzén, Temperature-dependent Hall effect measurements in low- compensated p-type 4H-SiC, Mater. Sci. Forum 457–460 (2004) 677–680, https://doi.org/10.4028/www.scientific.net/MSF.457-460.677.
[42] H. Matsuura, M. Komeda, S. Kagamihara, H. Iwata, R. Ishihara, T. Hatakeyama, T. Watanabe, K. Kojima, T. Shinohe, K. Arai, Dependence of acceptor levels and hole mobility on acceptor density and temperature in Al-doped p-type 4H-SiC epilayers, J. Appl. Phys. 96 (2004) 2708–2715, https://doi.org/10.1063/1.1775298.
[43] F. Schmid, M. Krieger, M. Laube, G. Pensl, G. Wagner, Silicon carbide: recent major advances, in: W.J. Choyke, H. Matsunami, G. Pensl (Eds.), Advances Text in Physics, Springer, Berlin, Germany, 2004517.
[44] A. Koizumi, J. Suda, T. Kimoto, Temperature and doping dependencies of electrical properties in Al-doped 4H-SiC epitaxial layers, J. Appl. Phys. 106 (2009) 013716–013718, https://doi.org/10.1063/1.3158565.
[45] J. Pernot, S. Contreras, J. Camassel, Electrical transport properties of aluminum-implanted 4H-SiC, J. Appl. Phys. 98 (2005) 023706–023709, https://doi.org/10.1063/1.1978987.
[46] M. Nawaz, On the assessment of few design proposals for 4H-SiC BJTs, Microelectr. J. 41 (2010) 801–808, https://doi.org/10.1016/j.mejo.2010.06.016.
[47] D. Stefanakis, K. Zekentes, TCAD models of the temperature and doping dependence of the bandgap and low field carrier mobility in 4H-SiC, Microelectron. Eng. 116 (2014) 65–71, https://doi.org/10.1016/j.mee.2013.10.002.